\documentclass[aps,pra,twocolumn,showpacs,floatfix,superscriptaddress]{revtex4}
\usepackage{graphicx}
\usepackage{dcolumn}
\usepackage{amsmath}
\usepackage{amssymb}
\usepackage{dsfont}
\usepackage{bm}
\usepackage{epstopdf}
\usepackage{multirow}
\newcommand{\fig}[1]{Figure~\ref{#1}}
\newcommand{\tab}[1]{Table~\ref{#1}}

\begin{document}

\title{Recurrence properties of unbiased coined quantum walks on infinite $d$ dimensional lattices}

\author{M. \v Stefa\v n\'ak\email[correspondence to:]{martin.stefanak@fjfi.cvut.cz}}
\affiliation{Department of Physics, FJFI \v CVUT, B\v
rehov\'a 7, 115 19 Praha 1 - Star\'e M\v{e}sto, Czech Republic}

\author{T. Kiss}
\affiliation{Department of Nonlinear and Quantum Optics, Research
Institute for Solid State Physics and Optics, Hungarian Academy of
Sciences, Konkoly-Thege u.29-33, H-1121 Budapest, Hungary}

\author{I. Jex}
\affiliation{Department of Physics, FJFI \v CVUT, B\v
rehov\'a 7, 115 19 Praha 1 - Star\'e M\v{e}sto, Czech Republic}


\date{\today}

\begin{abstract}
The P\'olya number characterizes the recurrence of a random walk. We apply the generalization of this concept to quantum walks [M. \v Stefa\v n\'ak, I. Jex and T. Kiss, Phys. Rev. Lett. \textbf{100}, 020501 (2008)] which is based on a specific measurement scheme. The P\'olya number of a quantum walk depends in general on the choice of the coin and the initial coin state, in contrast to classical random walks where the lattice dimension uniquely determines it. We analyze several examples to depict the variety of possible recurrence properties.
First, we show that for the class of quantum walks driven by tensor-product coins, the P\'olya number is independent of the initial conditions and the actual coin operators, thus resembling the property of the classical walks. We provide an estimation of the P\'olya number for this class of quantum walks. Second, we examine the 2-D Grover walk, which exhibits localisation and thus is recurrent, except for a particular initial state for which the walk is transient. We generalize the Grover walk to show that one can construct in arbitrary dimensions a quantum walk which is recurrent. This is in great contrast with classical walks which are recurrent only for the dimensions $d=1,2$. Finally, we analyze the recurrence of the 2-D Fourier walk. This quantum walk is recurrent except for a two-dimensional subspace of the initial states. We provide an estimation of the P\'olya number in its dependence on the initial state.
\end{abstract}

\maketitle


\section{Introduction}
\label{sec1}

Random walks (RWs) present a very useful tool in many branches of science \cite{overview}. Among others, the RW is one of the cornerstones of theoretical computer science \cite{rw:compsc1,rw:compsc2}. Indeed, it can be employed for algorithmic purposes solving problems such as graph connectivity \cite{graph:connect}, n-SAT \cite{3-sat} or approximating the permanent of a matrix \cite{matrix:perm}.

Quantum random walks (QWs) as a generalization of classical random walks to quantum systems have been proposed by Aharonov, Davidovich and Zagury \cite{aharonov}. The unitary time evolution can be considered either discrete as introduced by Meyer \cite{meyer1,meyer2} and Watrous \cite{watrous} leading to coined QWs or continuous as introduced by Farhi and Gutman \cite{farhi,childs}. Scattering quantum walks \cite{hillery:2003,hillery:2004,kosik:2005,hillery:2007} were proposed by Hillery, Bergou and Feldman as a natural generalization of coined QWs based on an interferometric analogy. The connection between the coined QW and the continuous time QW has been established recently \cite{strauch}.

The coined QW is well suited as an algorithmic tool \cite{kempe,ambainis}. Several algorithms based on coined QWs showing speed up over classical algorithms have been proposed \cite{shenvi:2003,ambainis:2003,kendon:2006,aurel:2007,magniez}. Various properties of coined QWs have been analyzed, e.g. the effects of the coin and the initial state \cite{2dw1,chandrashekar:2007,miyazaki}, absorbing barriers \cite{bach:2004}, the hitting times \cite{kempe:2005,krovi:2006a,krovi:2006b} or the effect of decoherence \cite{aurel:2007,kendon:2006b}. Hitting times for continuous QWs related to the quantum Zeno effect were considered in \cite{varbanov:2008}. Great attention has also been paid to the asymptotics of QWs \cite{nayak,carteret,Grimmett,konno:2002,konno:2005b}. In particular, localisation was found in 2-D QWs \cite{2dqw,2dw1,localisation} and in 1-D for a generalized QW \cite{1dloc,sato:2008}. Several experimental schemes have been proposed to realize coined QWs including cavity QED \cite{sanders}, linear optics \cite{jeong,pathak}, optical lattices \cite{eckert,dur} to BEC \cite{chandrashekar:2006}. By now, quantum walks form a well established part of quantum information theory \cite{bruss:leuchs}.

An interesting question for classical random walks is whether the particle eventually returns to the starting point of the random walk. The recurrence probability is known as the P\'olya number, after G. P\'olya who first discussed this property in classical RWs on infinite lattices in 1921 \cite{polya}. P\'olya pointed out the fundamental difference between walks in different dimensions. In three or higher dimensions the recurrence has a finite, non-unit probability depending exclusively on the dimension, whereas for walks in one or two dimensions the P\'olya number equals one. As a consequence, in three and higher dimensions the particle has a non-zero probability of escape \cite{hughes,domb:1954}. Recurrence in classical RWs is closely related to first passage times \cite{montroll:1964}. Recurrence in semi-infinite and finite QWs on lattices has been calculated in \cite{ambainis,yang:2007}.

In a recent letter \cite{prl} we have defined the P\'olya number for quantum walks on a $d$ dimensional lattice by extending the concept of recurrence. In this paper we calculate the P\'olya number for various coined QWs in one and two dimensions and construct arbitrary dimensional QWs exhibiting highly non-classical features.

Our paper is organized as follows: In Section~\ref{sec2} we review the concept of recurrence and the P\'olya number of random walks and its extension to quantum walks as defined in \cite{prl}. Both classical and quantum definitions lead to a similar criterion for recurrence determined by the asymptotic scaling of the return probability $p_0(t)$, as we show in Appendix~\ref{app:a}.

We dedicate Section~\ref{sec3} to the study of the asymptotic behaviour of the return probability $p_0(t)$. For this purpose we apply the Fourier transformation and the method of stationary phase. In particular, we demonstrate that the asymptotic behaviour of the return probability $p_0(t)$ is influenced by three factors: the topology of the walk, the choice of the coin operator and the initial coin state. Consequently, the nature of the QW for a fixed dimension can change from recurrent to transient and the actual value of the P\'olya number varies. This is in great contrast to classical random walks where the recurrence is uniquely determined by the dimension of the walk.

We use the results derived in Section~\ref{sec3} to determine the recurrence properties of several types of QWs. In Section~\ref{sec41} we treat unbiased 1-D QWs and show that all of them are recurrent independent of the coin operator or the initial coin state. We then generalize 1-D QWs in Section~\ref{sec42} to $d$ dimensions by considering an independent coin for each spatial dimension. We find that for this class of QWs the $p_0(t)$ is independent of the initial coin state and the actual form of the coin operator. Hence, a unique P\'olya number can be assigned to this class of QWs for each dimension $d$. In contrast with the classical RWs this class of QWs is recurrent only for $d=1$.

In Section~\ref{sec51} we analyze the recurrence of the 2-D Grover walk. This QW exhibits localisation \cite{localisation} and therefore is recurrent. However, for a particular initial state localisation disappears and the QW is transient. We find an approximation of the P\'olya number for this particular initial state. In Section~\ref{sec52} we employ the 2-D Grover walk to construct for arbitrary dimension $d$ a QW which is recurrent. This is in great contrast with the classical RWs, which are recurrent only for the dimensions $d=1,2$.

Finally, in Section \ref{sec6} we analyze the 2-D Fourier walk. This QW is recurrent except for a two-parameter family of initial states for which it is transient. For the latter case we find an approximation of the P\'olya number depending on the parameters of the initial state.

We conclude by presenting an outlook in Section~\ref{sec7}.

\section{Recurrence and the P\'olya number of random walks}
\label{sec2}

Random walks are classically defined as the probabilistic evolution of the position of a point-like particle on a discrete graph. Starting the walker from a well-defined graph point (the origin) one can ask about the probability that the walker returns there at least once during the time evolution. The event that the walker is not present at the origin at any time instant is just the complement of the event corresponding to recurrence. The probability of the latter is called the P\'olya number. Classical random walks are said to be {\it recurrent} or  {\it transient} depending on whether their P\'olya number equals to one, or is less than one, respectively.

The P\'olya number of a classical random walk can be defined in the following way \cite{revesz}. Let $q_0(t)$ be the probability that the walker returns to the origin for the {\it first time} after $t$ steps. Since these events are mutually exclusive we can add up their probabilities and the series
\begin{equation}
P\equiv\sum\limits_{t=1}^\infty q_0(t)
\label{polya:1}
\end{equation}
is the probability that at least once the particle has returned to the origin, i.e. the P\'olya number. However, the definition (\ref{polya:1}) is not very practical for determining the recurrence nature of a random walk. Nevertheless, we can express the P\'olya number in terms of the probability $p_0(t)$ that the particle is at the origin at any given time instant $t$. Indeed, it is straightforward to show that
\begin{equation}
P = 1-\frac{1}{\sum\limits_{t=0}^{+\infty}p_0(t)}.
\label{pol:cl}
\end{equation}
Hence, the recurrence behaviour of a RW is determined solely by the infinite sum
\begin{equation}
{\cal S} \equiv \sum_{t=0}^{\infty}p_0(t).
\label{sum}
\end{equation}
We find that $P$ equals unity if and only if the series ${\cal S}$ diverges \cite{revesz}. In such a case the walk is recurrent. On the other hand, if the series $\cal S$ converges, the P\'olya number $P$ is strictly less than unity and the walk is transient. We will restrict the considered graphs to $d$ dimensional uniform infinite square lattices and the considered walks to balanced ones, {\it i.e.} the probability to take one step is equal in each of the possible directions along the lattice. P\'olya proved in \cite{polya} that in one and two dimensions, such RWs are recurrent while for $d>2$ they are transient.

Discrete time coined quantum walks are generalizations of classical random walks. Here the dynamics is a unitary evolution on a composite Hilbert space consisting of the external positions on the graph and the internal state determining the direction of the following step. For quantum walks one would like to keep the same definition of the P\'olya number, i.e. the probability of returning to the origin at least once during the time evolution. However, to be able to talk about the position of a particle in quantum mechanics one must specify when and which type of measurement is performed. We have given a definition of the P\'olya number for QWs \cite{prl} by specifying the following measurement scheme. From an ensemble of identically prepared QW systems take one, let it evolve for one step, measure the position and discard this system; take a second, identically prepared system, let it evolve for two steps, measure the position and then discard it; continue until it is found at the origin.
In the $t$-th trial we do not find the particle at the origin with the probability $1-p_0(t)$. Since the individual trials are independent the product
\begin{equation}
\overline{P}_n = \prod_{t=1}^n(1-p_0(t))
\end{equation}
gives the probability that we have not found the particle at the origin in the first $n$ trials. In the complementary event which occurs with the probability
\begin{equation}
P_n = 1-\prod_{t=1}^n(1-p_0(t))
\label{polya:approx}
\end{equation}
the particle was found at least once at the origin, by which we understand the recurrence. Hence, we define the P\'olya number of a QW by extending $n$ to infinity
\begin{equation}
P = 1-\prod\limits_{t=1}^{+\infty}(1-p_0(t)).
\label{polya:def}
\end{equation}
As we show in Appendix~\ref{app:a}, Eq. (\ref{polya:def}) leads to the same criterion for recurrence in terms of $p_0(t)$ --- the infinite product in Eq. (\ref{polya:def}) vanishes if and only if the series $\cal S$ diverges \cite{jarnik}. In such a case the P\'olya number of a QW is unity and we call such QWs recurrent. If the series $\cal S$ converges, then the product in Eq. (\ref{polya:def}) does not vanish and the P\'olya number of a QW is less than one. In accordance with the classical terminology we call such QWs transient. We use the formula in Eq. (\ref{polya:approx}) to estimate the P\'olya number of a QW.

To conclude this section, the recurrence nature of quantum random walks is determined by the decay of the return probability $p_0(t)$, in the same way as for classical random walks.


\section{Asymptotics of quantum walks}
\label{sec3}

Before we turn to the recurrence of QWs we analyze the asymptotics of the return probability $p_0(t)$. Our approach is based on the Fourier transformation and the method of stationary phase. We consider random walks where the walker has to leave its actual position at each step. Hence, $p_0(t)\equiv 0$ for odd times and it is sufficient to consider only even times $2t$. For simplicity we omit the factor of 2 from now on.

\subsection{Description of quantum walks on $\mathds{Z}^d$}
\label{sec31}

We consider quantum walks on an infinite $d$ dimensional lattice $\mathds{Z}^d$. The Hilbert space of the quantum walk can be written as a tensor product
\begin{equation}
\mathcal{H} = \mathcal{H}_P\otimes\mathcal{H}_C
\end{equation}
of the position space
\begin{equation}
\mathcal{H}_P=\ell^2(\mathds{Z}^d)
\end{equation}
and the coin space $\mathcal{H}_C$. The position space is spanned by the vectors $|\textbf{m}\rangle$ corresponding to the walker being at the lattice point $\textbf{m}$, i.e.
\begin{equation}
\mathcal{H}_P=\text{Span}\left\{|\textbf{m}\rangle|\quad\textbf{m}=\left\{m_1,\ldots,m_d\right\}\in\mathds{Z}^d\right\}.
\end{equation}
The coin space $\mathcal{H}_C$ is determined by the topology of the walk. In particular, its dimension $c$ is given by the number of possible displacements in a single step. We denote the displacements by vectors
\begin{equation}
\label{shift}
\mathbf{e}_i\in\mathds{Z}^d,\quad i=1,\ldots,c.
\end{equation}
Hence, the walker can move from $\textbf{m}$ to any of the points $\textbf{m}+\textbf{e}_i, i=1,\ldots,c$ in a single step. We restrict ourselves to unbiased walks where the displacements satisfy the condition
\begin{equation}
\sum_{i=1}^c\mathbf{e}_i=\mathbf{0}.
\label{sum:shift:0}
\end{equation}
We define an orthonormal basis in the coin space by assigning to every displacement $\mathbf{e}_i$ the basis vector $|\mathbf{e}_i\rangle$, i.e.
\begin{equation}
\mathcal{H}_C = \textrm{Span}\left\{|\mathbf{e}_i\rangle|i=1,\ldots,c\right\}.
\end{equation}
A single step of the QW is given by
\begin{equation}
U=S \left(I_P\otimes C\right).
\label{qw:time}
\end{equation}
Here $I_P$ denotes the unit operator acting on the position space $\mathcal{H}_P$. The coin flip operator $C$ is applied on the coin state before the displacement $S$ itself. The coin flip $C$ can be in general an arbitrary unitary operator acting on the coin space $\mathcal{H}_C$. We restrict ourselves to unbiased walks for which the coin $C$ meets the requirement
\begin{equation}
\left|C_{ij}\right|\equiv\left|\langle\mathbf{e}_i|C|\mathbf{e}_j\rangle\right|=\frac{1}{\sqrt{c}},
\end{equation}
i.e. all matrix elements of $C$ must have the same absolute value. Such matrices are closely related to the Hadamard matrices \cite{dita}.

The displacement itself is represented by the step operator $S$
\begin{equation}
S = \sum\limits_{\mathbf{m},i}|\mathbf{m}+\mathbf{e}_i\rangle\langle\mathbf{m}|\otimes|\mathbf{e}_i\rangle\langle\mathbf{e}_i|,
\end{equation}
which moves the walker from the site $\mathbf{m}$ to $\mathbf{m}+\mathbf{e}_i$ if the state of the coin is $|\mathbf{e}_i\rangle$.

Let the initial state of the walker be
\begin{equation}
|\psi(0)\rangle \equiv \sum\limits_{\mathbf{m},i}\psi_i(\mathbf{m},0)|\mathbf{m}\rangle\otimes|\mathbf{e}_i\rangle.
\end{equation}
Here $\psi_i(\mathbf{m},0)$ is the probability amplitude of finding the walker at time $t=0$ at the position $\mathbf{m}$ in the coin state $|\mathbf{e}_i\rangle$. The state of the walker after $t$ steps is given by successive application of the time evolution operator given by Eq. (\ref{qw:time}) on the initial state
\begin{equation}
|\psi(t)\rangle \equiv \sum\limits_{\mathbf{m},i}\psi_i(\mathbf{m},t)|\mathbf{m}\rangle\otimes|\mathbf{e}_i\rangle=U^t|\psi(0)\rangle.
\label{time:evol}
\end{equation}
The probability of finding the walker at the position $\textbf{m}$ at time $t$ is given by the summation over the coin state, i.e.
\begin{eqnarray}
\nonumber p(\textbf{m},t) & \equiv & \sum_{i=1}^c|\langle\textbf{m}|\langle\mathbf{e}_i|\psi(t)\rangle|^2 = \sum_{i=1}^c|\psi_i(\mathbf{m},t)|^2\\
 & = & ||\psi(\textbf{m},t)||^2.
\end{eqnarray}
Here we have introduced $c$ component vectors
\begin{equation}
\psi(\mathbf{m},t)\equiv{\left(\psi_1(\mathbf{m},t),\psi_2(\mathbf{m},t),\ldots,\psi_c(\mathbf{m},t)\right)}^T
\label{prob:ampl}
\end{equation}
of probability amplitudes. We rewrite the time evolution equation (\ref{time:evol}) for the state vector $|\psi(t)\rangle$ into a set of difference equations
\begin{equation}
\psi(\mathbf{m},t) = \sum_l C_l\psi(\mathbf{m}-\mathbf{e}_l,t-1)
\label{time:evol2}
\end{equation}
for probability amplitudes $\psi(\mathbf{m},t)$. Here the matrices $C_l$ have all entries equal to zero except for the $l$-th row which follows from the coin-flip operator $C$, i.e.
\begin{equation}
\langle\mathbf{e}_i\left|C_l\right|\mathbf{e}_j\rangle = \delta_{il}\langle\mathbf{e}_i\left|C\right|\mathbf{e}_j\rangle.
\end{equation}

\subsection{Solution via Fourier Transformation}
\label{sec32}

The QWs we consider are translationally invariant which manifests itself in the fact that the matrices $C_l$ on the right-hand side of Eq. (\ref{time:evol2}) are independent of $\mathbf{m}$. Hence, the time evolution equations (\ref{time:evol2}) simplify considerably with the help of the Fourier transformation
\begin{equation}
\tilde{\psi}(\mathbf{k},t)\equiv\sum\limits_\mathbf{m}\psi(\mathbf{m},t) e^{i \mathbf{m}\cdot\mathbf{k}}, \quad \mathbf{k}\in\mathbb{K}^d.
\label{qw:ft}
\end{equation}
The Fourier transformation defined by Eq. (\ref{qw:ft}) is an isometry between $\ell^2(\mathds{Z}^d)$ and $L^2(\mathds{K}^d)$ where $\mathds{K}=(-\pi,\pi]$ can be thought of as the phase of a unit circle in $\mathds{R}^2$.

The time evolution in the Fourier picture turns into a single difference equation
\begin{equation}
\tilde{\psi}(\mathbf{k},t)=\widetilde{U}(\mathbf{k})\tilde{\psi}(\mathbf{k},t-1).
\label{qw:te:fourier}
\end{equation}
Here we have introduced the time evolution operator in the Fourier picture
\begin{eqnarray}
\nonumber \widetilde{U}(\mathbf{k}) & \equiv & D(\mathbf{k}) C\\
D(\mathbf{k}) & \equiv & \textrm{diag}\left(e^{-i\mathbf{e}_1\cdot\mathbf{k}},\ldots,e^{-i\mathbf{e}_c\cdot\mathbf{k}}\right).
\label{teopF}
\end{eqnarray}
We find that $\widetilde{U}(\mathbf{k})$ is determined both by the coin $C$ and the topology of the QW through the diagonal matrix $D(\mathbf{k})$ containing the displacements $\mathbf{e}_i$.

We solve the difference equation (\ref{qw:te:fourier}) by formally diagonalising the matrix $\widetilde{U}(\mathbf{k})$. Since it is a unitary matrix its eigenvalues can be written in the form
\begin{equation}
\lambda_j(\mathbf{k})=\exp{\left(i \omega_j(\mathbf{k})\right)}.
\label{eigen:C}
\end{equation}
We denote the corresponding eigenvectors as $v_j(\mathbf{k})$. Using this notation the state of the walker in the Fourier picture at time $t$ reads
\begin{equation}
\tilde{\psi}(\mathbf{k},t) = \sum_j e^{i\omega_j(\mathbf{k})t}\left(\tilde{\psi}(\mathbf{k},0),v_j(\mathbf{k})\right)v_j(\mathbf{k}),
\label{sol:k}
\end{equation}
where $\left(\ ,\ \right)$ denotes the scalar product in the $c$ dimensional space. Finally, we perform the inverse Fourier transformation and find the exact expression for the probability amplitudes
\begin{equation}
\psi(\mathbf{m},t) = \int_{\mathds{K}^d}\frac{d\mathbf{k}}{(2\pi)^d}\ \widetilde{\psi}(\mathbf{k},t)\ e^{-i \mathbf{m}\cdot\mathbf{k}}
\label{inv:f}
\end{equation}
in the position representation.

We are interested in the recurrence nature of QWs. As we have shown in Section~\ref{sec2} the recurrence of a QW is determined by the asymptotic behaviour of the probability that the walker returns to the origin at time $t$
\begin{equation}
p_0(t)\equiv p(\mathbf{0},t)=\left\|\psi(\mathbf{0},t)\right\|^2.
\label{po}
\end{equation}
Hence, we set $\mathbf{m}=\mathbf{0}$ in Eq. (\ref{inv:f}). Moreover, in analogy with the classical problem of P\'olya we restrict ourselves to QWs which start at origin. Hence, the initial condition reads
\begin{equation}
\psi(\mathbf{m},0)=\delta_{\mathbf{m},\mathbf{0}}\psi, \quad \psi\equiv\psi(\mathbf{0},0)
\label{init:cond}
\end{equation}
and its Fourier transformation $\tilde{\psi}(\mathbf{k},0)$ entering Eq. (\ref{sol:k}) is identical to the initial state of the coin
\begin{equation}
\tilde{\psi}(\mathbf{k},0)=\psi,
\end{equation}
which is a $c$-component vector. We note that due to the Kronecker delta in Eq. (\ref{init:cond}) the Fourier transformation $\tilde{\psi}(\mathbf{k},0)$ is a constant vector.

Using the above assumptions we find the exact expression for the return probability
\begin{equation}
p_0(t) = \left|\sum_{j=1}^c I_j(t)\right|^2
\end{equation}
where $I_j(t)$ are given by the integrals
\begin{eqnarray}
\nonumber I_j(t) & = & \int\limits_{\mathbb{K}^d}\frac{d\mathbf{k}}{(2\pi)^d}\ e^{i\omega_j(\mathbf{k})t}\ f_j(\mathbf{k})\\
f_j(\mathbf{k}) & = & \left(\psi,v_j(\mathbf{k})\right)\ v_j(\mathbf{k}).
\label{psi:0}
\end{eqnarray}

\subsection{Asymptotics of $p_0(t)$ via the method of stationary phase}
\label{sec33}

Let us now discuss how the additional freedom we have at hand for QWs influences the asymptotics of the return probability $p_0(t)$. For simplicity we suppose that the asymptotic behaviour of $p_0(t)$ arises from $|I_j(t)|^2$. We suppose that the functions $\omega_j(\mathbf{k})$ and  $f_j(\mathbf{k})$ entering $I_j(t)$ are smooth. According to the method of stationary phase \cite{statphase} the major contribution to the integral $I_j(t)$ comes from the saddle points $\mathbf{k}^0$ of the phase $\omega_j(\mathbf{k})$, i.e. by the points where the gradient of the phase vanishes
\begin{equation}
\left.\vec{\nabla}\omega_j(\mathbf{k})\right|_{\mathbf{k}=\mathbf{k}^0} = \mathbf{0}.
\end{equation}
The asymptotic behaviour of $I_j(t)$ is then determined by the saddle point with the greatest degeneracy given by the dimension of the kernel of the Hessian matrix
\begin{equation}
H^{(j)}_{m,n}(\mathbf{k})\equiv \frac{\partial^2 \omega_j(\mathbf{k})}{\partial k_m\partial k_n}
\end{equation}
evaluated at the saddle point. The function $f_j(\mathbf{k})$ entering the integral $I_j(t)$ determines only the pre-factor of the asymptotic behaviour. We now discuss how does the existence, configuration and number of saddle points affect the asymptotic behaviour of $I_j(t)$. As a rule of thumb, the decay of the return probability $p_0(t)$ can slow down with the increase in the number of saddle points. We briefly summarize the cases of no saddle points, finite number of saddle points and a continuum of saddle points.

({\it i}) No saddle points

If $\omega_j(\mathbf{k})$ has no saddle points then $I_j(t)$ decays faster than any inverse polynomial in $t$. Consequently, the decay of the return probability $p_0(t)$ would also be exponential. However, among the examples we have considered such a situation was not found.

({\it ii}) Finite number of saddle points

If $\omega_j(\mathbf{k})$ has a finite number of non-degenerate saddle points, i.e. the determinant of the Hessian matrix $H$ is non-zero for all saddle points, and the function $f_j(\mathbf{k})$ does not vanish at the saddle points then the contribution from all saddle points to the integral $I_j(t)$ is of the order $t^{-d/2}$ and $p_0(t) \sim t^{-d}$.  Though the contributions from the distinct saddle points might have different relative phases and can interfere destructively we have never encountered a complete cancelation of all contributions.

({\it iii}) Continuum of saddle points

If $\omega_j(\mathbf{k})$ has a continuum of saddle points then the dimension of the continuum determines the decay of the integral $I_j(t)$. The case of 2-D integrals with curves of stationary points are treated in \cite{statphase}. It is shown that the contribution from the continuum of stationary points to the integral $I_j(t)$ is of the order $t^{-1/2}$. This is greater than the contribution arising from a discrete saddle point which is of the order $t^{-1}$. Hence, the continuum of saddle points has effectively slowed-down the decay of the integral $I_j(t)$. Consequently, the leading order term of the return probability is $p_0(t)\sim t^{-1}$. Similar results can be expected for higher dimensional QWs where the phase $\omega_j(\mathbf{k})$ has a continuum of saddle points. A special case for a continuum of saddle points is when $\omega_j(\mathbf{k})$ does not depend on $n$ variables but it has a finite number of saddle points with respect to the remaining $d-n$ variables. Indeed, such an $\omega_j(\mathbf{k})$ has obviously a zero derivative with respect to those $n$ variables. If, in addition, $I_j(t)$ factorizes into the product of time-independent and time-dependent integrals over $n$ and $d-n$ variables and the time-independent integral does not vanish then $I_j(t)$ behaves asymptotically like $t^{-(d-n)/2}$. Hence, the asymptotic behaviour of the return probability is $p_0(t)\sim{t^{-(d-n)}}$. In the extreme case when the phase $\omega_j(\mathbf{k})$ does not depend on $\mathbf{k}$ at all we can extract the time dependence out of the integral $I_j(t)$. If the remaining time independent integral does not vanish then $p_0(t)$ is a non-zero constant.

In the above discussion we have assumed that the function $f_j(\mathbf{k})$ is non-vanishing for $\mathbf{k}$ values corresponding to the saddle points. However, the initial state $\psi$ can be orthogonal to the eigenvector $v_j(\mathbf{k})$ for $\mathbf{k}=\mathbf{k}^0$ corresponding to the saddle point. In such a case the function $f_j(\mathbf{k})$ vanishes for $\mathbf{k}=\mathbf{k}^0$ and the saddle point $\mathbf{k}^0$ does not contribute to the integral $I_j(t)$. Consequently, the decay of $p_0(t)$ can speed up.

In Section \ref{sec2} we have shown that the recurrence nature of the QWs is determined by convergence or divergence of the sum (\ref{sum}) which in turn depends on the speed of decay of the return probability $p_0(t)$. Hence, for QWs we might change the recurrence behaviour and the actual value of the P\'olya number by altering the initial state $\psi$, coin flip $C$ and the topology of the walk determined by the displacements $\mathbf{e}_i$.

In the following sections we make use of the above derived results and determine the recurrence behaviour and the P\'olya number of several types of QWs. We concentrate on the effect of the coin operators and the initial states. For this purpose we fix the topology of the walks. We consider QWs where the displacements $\mathbf{e}_i$ have all entries equal to $\pm 1$
\begin{equation}
\mathbf{e}_1 = \left(1,\ldots,1\right)^T,\ldots, \mathbf{e}_{2^d} = \left(-1,\ldots,-1\right)^T.
\end{equation}
In such a case the coin space has the dimension $c=2^d$ and the diagonal matrix $D(\mathbf{k})$ can be written as a tensor product
\begin{equation}
D(\textbf{k}) = D(k_1)\otimes\ldots\otimes D(k_d)
\label{dk2}
\end{equation}
of $2\times 2$ diagonal matrices $D(k_j)=\textrm{diag}(e^{-ik_j},e^{ik_j})$. This fact greatly simplifies the diagonalisation of the time evolution operator in the Fourier picture $\widetilde{U}(\mathbf{k})$.


\section{Recurrence of 1-D QWs and QWs with tensor-product coins}
\label{sec4}

We begin this section with the analysis of unbiased 1-D QWs. We find that all unbiased 1-D QWs are recurrent independently of the initial coin state and the actual form of the coin operator. We then generalize unbiased 1-D QWs to $d$ dimensions by considering coins which can be written as a tensor products of $d$ $2\times 2$ matrices, i.e. we consider independent coin for each spatial dimension. This class of $d$ dimensional QWs maintains some properties of the 1-D QWs. In particular, the asymptotic behaviour of the probability $p_0(t)$ is independent of the initial coin state and the actual form of the coin operator. Hence, a unique P\'olya number can be assigned to this class of QWs for each dimension $d$. In contrast with the classical RWs they are recurrent only for $d=1$.

\subsection{Unbiased QWs in one dimension}
\label{sec41}

Let us start with the analysis of the recurrence behaviour of unbiased 1-D QWs. The general form of the unbiased coin for 1-D quantum walk is given by
\begin{equation}
C(\alpha,\beta) = \frac{1}{\sqrt{2}}\left(
        \begin{array}{rr}
          e^{i\alpha} & e^{-i\beta} \\
          e^{i\beta} & -e^{-i\alpha} \\
        \end{array}
      \right).
\label{1d:unbiased}
\end{equation}
We find that the time evolution operator in the Fourier picture
\begin{equation}
\widetilde{U}(k,\alpha,\beta) = D(k) C(\alpha,\beta)
\label{1d:Had}
\end{equation}
has eigenvalues $e^{i\omega_i(k,\alpha)}$ with the phases $\omega_i(k,\alpha)$ given by
\begin{equation}
\sin\omega_1(k,\alpha)=-\frac{\sin{(k-\alpha)}}{\sqrt{2}}, \quad \omega_2(k,\alpha)=\pi-\omega_1(k,\alpha).
\label{phase:1d}
\end{equation}
Thus the derivatives of $\omega_i$ with respect to $k$ reads
\begin{equation}
\frac{d\omega_1(k,\alpha)}{dk}=-\frac{d\omega_2(k,\alpha)}{dk}=-\frac{\cos{(k-\alpha)}}{\sqrt{2-\sin^2{(k-\alpha)}}}
\label{der:1d}
\end{equation}
and we find that the phases $\omega_i(k,\alpha)$ have common non-degenerate saddle points $k^0=\alpha\pm\pi/2$. It follows that $p_0(t)$ behaves asymptotically like $t^{-1}$, independently of the coin parameters $\alpha,\beta$. Moreover, the asymptotic behaviour is independent of the initial state. Indeed, no non-zero initial state $\psi$ exists which is orthogonal to both eigenvectors at the common saddle points $k^0=\alpha\pm\pi/2$. Hence, all unbiased 1-D quantum walks are recurrent, i.e. the P\'olya number equals one, independently of the initial coin state and the coin. However, none of the QWs from the class described by Eq. (\ref{1d:unbiased}) exhibits localisation, since for all of them the probability $p_0(t)$ converges to zero. We note that one can achieve localisation in 1-D by considering generalized QWs for which the coin has more degrees of freedom \cite{1dloc}.

\subsection{Higher dimensional QWs with tensor-product coins}
\label{sec42}

We now turn to the class of QWs with independent coin for each spatial dimension, i.e. the coin flip operator has the form of the tensor product of $d$ $2\times 2$ matrices
\begin{equation}
C^{(d)}(\bm{\alpha},\bm{\beta}) = C(\alpha_1,\beta_1)\otimes\ldots\otimes C(\alpha_d,\beta_d).
\label{C:ind:x}
\end{equation}
It follows that also the time evolution operator in the Fourier picture described by Eq. (\ref{teopF}) has the form of the tensor product
\begin{equation}
\widetilde{U}^{(d)}(\textbf{k},\bm{\alpha},\bm{\beta}) = \widetilde{U}(k_1,\alpha_1,\beta_1)\otimes\ldots\otimes \widetilde{U}(k_d,\alpha_d,\beta_d)
\label{C:ind}
\end{equation}
of $d$ 1-D time evolution operators given by Eq. (\ref{1d:Had}) with different parameters $k_i,\alpha_i,\beta_i$. Hence, the phases of the eigenvalues of the matrix $\widetilde{U}^{(d)}(\textbf{k},\bm{\alpha},\bm{\beta})$ have the form of the sum
\begin{equation}
\omega_j(\textbf{k},\bm{\alpha})=\sum_{l=1}^d \omega_{j_l}(k_l,\alpha_l).
\label{eigenval:Cind}
\end{equation}
Therefore we find that the asymptotic behaviour of this class of QWs follows directly from the asymptotics of the 1-D QWs. Indeed, the derivative of the phase $\omega_j(\textbf{k},\bm{\alpha})$ with respect to $k_l$ reads
\begin{equation}
\frac{\partial \omega_j(\textbf{k},\bm{\alpha})}{\partial k_l} = \frac{d \omega_{j_l}(k_l,\alpha_l)}{d k_l},
\label{der:phase:Cind}
\end{equation}
and so $\omega_j(\textbf{k},\bm{\alpha})$ has a saddle point $\textbf{k}^0=\left(k_1^0,k_2^0,\ldots,k_d^0\right)$ if and only if for all $l=1,\ldots,d$ the point $k_l^0$ is the saddle point of $\omega_{j_l}(k_l,\alpha_l)$. As we have found from Eq. (\ref{der:1d}) the saddle points of $\omega_{j_l}$ are $k^0_l=\alpha_l\pm\pi/2$. Hence, all phases $\omega_j(\textbf{k},\bm{\alpha})$ of the eigenvalues of the matrix $\widetilde{U}^{(d)}(\textbf{k},\bm{\alpha},\bm{\beta})$ have $2^d$ common saddle points $\textbf{k}^0=\left(\alpha_1\pm\pi/2,\ldots,\alpha_d\pm\pi/2\right)$. It follows that the asymptotic behaviour of the probability $p_0(t)$ is determined by
\begin{equation}
p_0^{(d)}(t)\sim t^{-d}.
\label{asymp:Cind}
\end{equation}
As follows from the results for 1-D QWs the asymptotic behaviour given by Eq. (\ref{asymp:Cind}) is independent of the initial coin state and of the coin parameters $\bm{\alpha},\bm{\beta}$. Compared to classical walks this is a quadratically faster decay of the probability $p_0(t)$ which is due to the quadratically faster spreading of the probability distribution of the 1-D QWs.

\begin{figure}
\begin{center}
\includegraphics[width=0.45\textwidth]{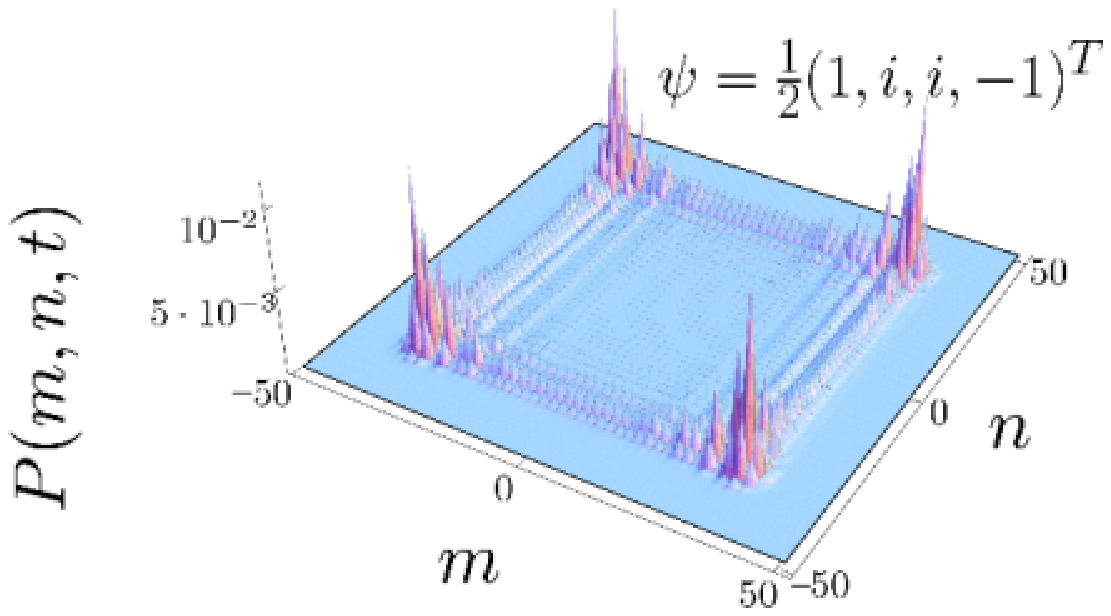}\vspace{12pt}
\includegraphics[width=0.45\textwidth]{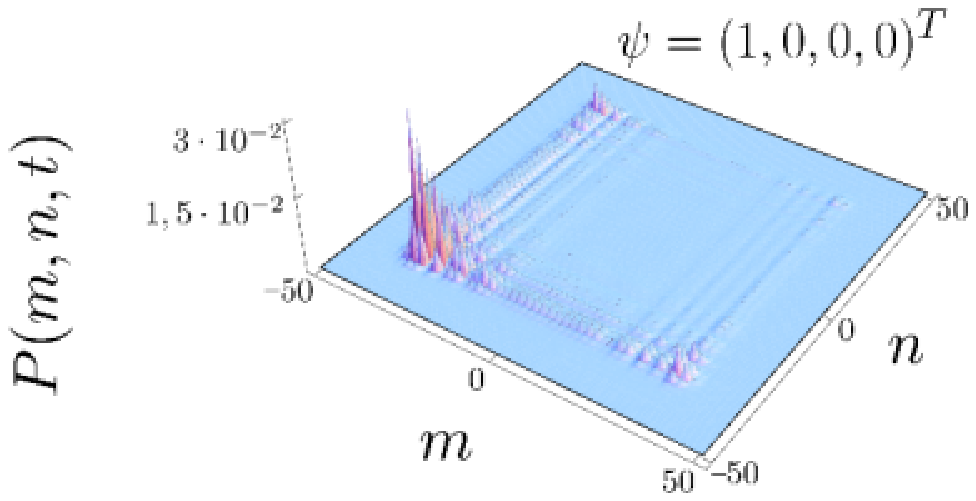}\vspace{12pt}
\includegraphics[width=0.37\textwidth]{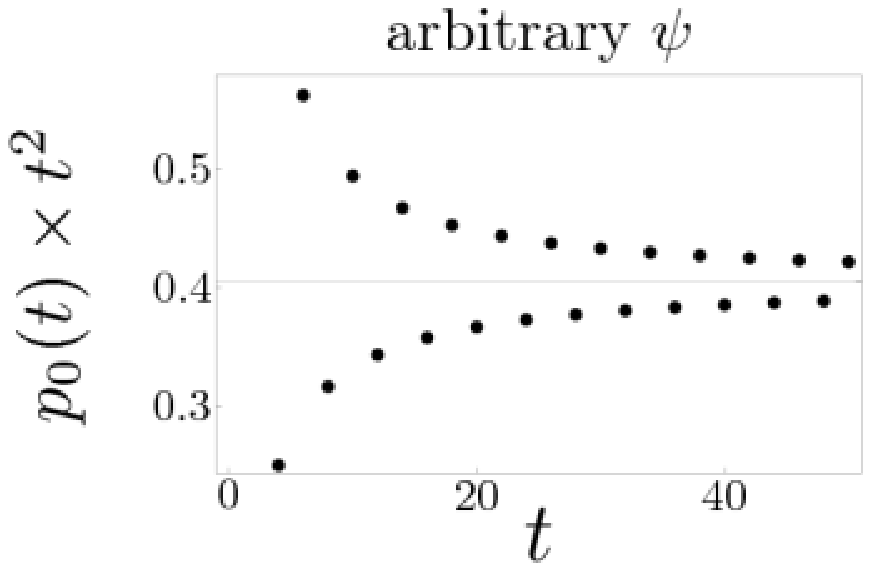}
\caption{Probability distribution of the 2-D Hadamard walk after 50 steps and the probability $p_0(t)$ for different choices of the initial state. In the upper plot we choose the initial state $\frac{1}{2}(1,i,i,-1)^T$ which leads to a symmetric probability distribution, whereas in the middle plot we choose the initial state $(1,0,0,0)^T$ resulting in a dominant peak of the probability distribution in the lower-left corner of the $(m,n)$ plane. However, the initial state influences the probability distribution only near the edges. The probability $p_0(t)$ is unaffected and is the same for all initial coin states. The lower plot confirms the asymptotic behaviour $p_0(t)\sim t^{-2}$ independent of the initial state.}
\label{had:2d}
\end{center}
\end{figure}

We illustrate the results for 2-D Hadamard walk driven by the coin which is a tensor product of two $2\times 2$ Hadamard matrices
\begin{equation}
C^{(2)}(\textbf{0},\textbf{0}) = \frac{1}{2}\left(
                      \begin{array}{rrrr}
                        1 & 1 & 1 & 1\\
                        1 & -1 & 1 & -1 \\
                        1 & 1 & -1 & -1\\
                        1 & -1 & -1 & 1\\
                      \end{array}
                    \right)
\end{equation}
in \fig{had:2d}. Here we show the probability distribution and the probability $p_0(t)$. The first row indicates that the initial state of the coin influences mainly the edges of the probability distribution. However, the probability $p_0(t)$ is unaffected and is exactly the same for all initial states. The lower plot confirms the asymptotic behaviour $p_0(t)\sim t^{-2}$.


Since the return probability $p_0(t)$ decays like $t^{-d}$ we find that $d$ dimensional quantum walks with tensor-product coins are recurrent only for dimension $d=1$ and are transient for all higher dimensions $d\geq 2$. Moreover, the whole sequence of probabilities $p_0(t)$ is independent of the initial state and the coin $C^{(d)}(\bm{\alpha},\bm{\beta})$. Hence, the P\'olya number for this class of QWs depends only on the dimension of the walk $d$, thus resembling the property of the classical walks. On the other hand, this class of QWs is transient for the dimension $d=2$ and higher. This is a direct consequence of the faster decay of the probability at the origin which, in this case, cannot be compensated for by interference.

Let us now estimate the value of the P\'olya number for the class of QWs with tensor-product coins and the dimension $d\geq 2$. As depicted in the lowest plot of \fig{had:2d} the return probability approaches quite rapidly its asymptotic
\begin{equation}
p_0^{(d)}(t)\approx\frac{1}{(\pi t)^{d}}.
\end{equation}
Hence, already first few terms of the product in Eq. (\ref{polya:approx}) are sufficient to estimate the value of the P\'olya number. Taking into account the first three terms of $p_0^{(d)}(t)$  which are found to be
\begin{equation}
p_0^{(d)}(2)=\frac{1}{2^d},\quad p_0^{(d)}(4)=p_0^{(d)}(6)=\frac{1}{8^d},
\end{equation}
we obtain the following approximation of the P\'olya number
\begin{equation}
P^{(d)}\approx 1 - \left(1-\frac{1}{2^d}\right)\left(1-\frac{1}{8^d}\right)^2.
\label{Polya:ind:est}
\end{equation}
We compare the estimation in Eq. (\ref{Polya:ind:est}) with the numerical results obtained from the simulation of the QWs with 1000 steps in the \tab{tab1} and find that they are in excellent agreement.

\begin{table}
\begin{center}
\begin{tabular}{|c|c|c|c|}
  \hline
  \multirow{2}{*}{Dimension} & \multirow{2}{*}{Simulation} & \multirow{2}{*}{Estimation (\ref{Polya:ind:est})} & \multirow{2}{*}{Error ($\%$)}\\
  & & & \\\hline
  \multirow{2}{*}{2} & \multirow{2}{*}{0.29325} & \multirow{2}{*}{0.27325} & \multirow{2}{*}{6.8}\\
  & & & \\\hline
  \multirow{2}{*}{3} & \multirow{2}{*}{0.12947} & \multirow{2}{*}{0.12841} & \multirow{2}{*}{0.82}\\
  & & & \\\hline
  \multirow{2}{*}{4} & \multirow{2}{*}{0.06302} & \multirow{2}{*}{0.06296} & \multirow{2}{*}{0.01}\\
  & & & \\\hline
  \multirow{2}{*}{5} & \multirow{2}{*}{0.031313} & \multirow{2}{*}{0.031309} & \multirow{2}{*}{0.01}\\
  & & & \\
  \hline
\end{tabular}
\caption{Comparison of the P\'olya numbers for the class of $d$ dimensional QWs with tensor product coins obtained from the numerical simulation and the and the analytical estimation of Eq. (\ref{Polya:ind:est}).}
\label{tab1}
\end{center}
\end{table}


\section{Recurrent quantum walks based on the 2-D Grover walk}
\label{sec5}

We now turn to the QWs based on the 2-D Grover walk. This QW has been extensively studied by many authors \cite{2dqw,2dw1,localisation,prl}. We re-derive the properties of this QW using the tools developed in Section~\ref{sec3}. We find that the 2-D Grover walk exhibits localisation and is therefore recurrent except for a particular initial state. We find an estimation of the P\'olya number in the latter case.

Employing the 2-D Grover walk we construct in arbitrary dimensions a QW which is recurrent, except for a subspace of initial states. This is in striking contrast to the classical random walks which are recurrent only for the dimensions $d=1,2$.

\subsection{2-D Grover walk}
\label{sec51}

\begin{figure}
\begin{center}
\includegraphics[width=0.45\textwidth]{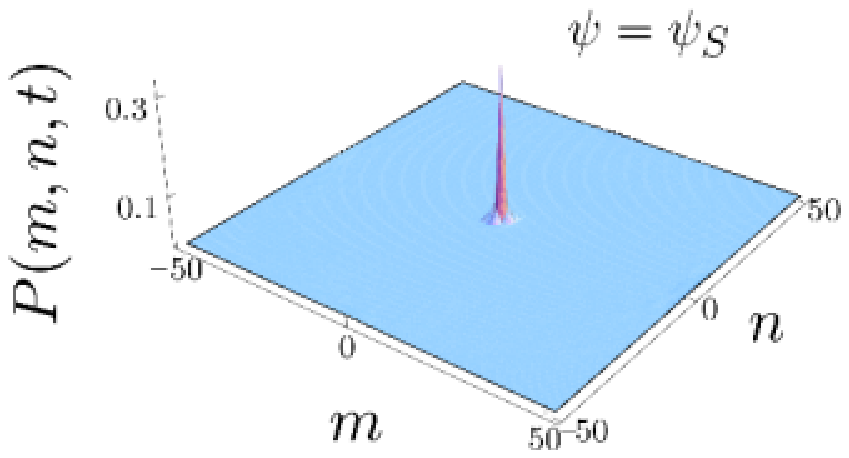}\vspace{12pt}
\includegraphics[width=0.45\textwidth]{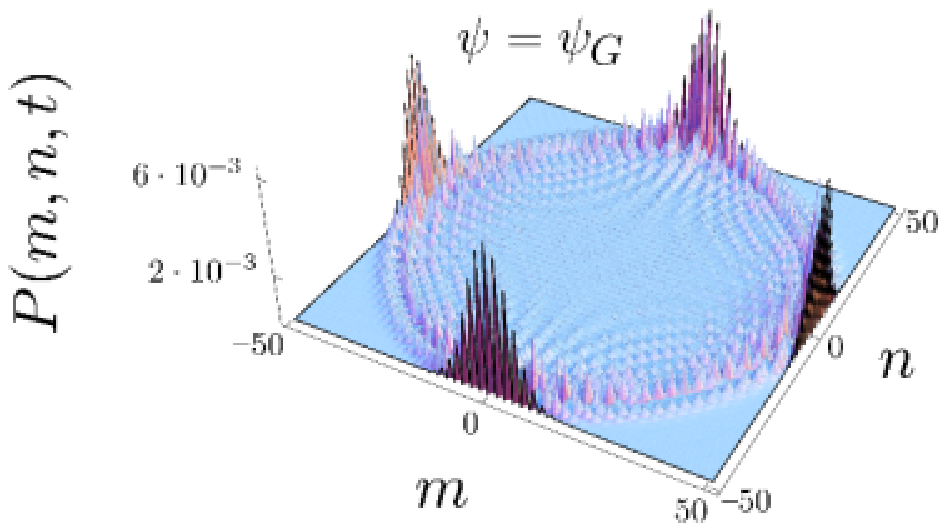}
\caption{Probability distribution of the Grover walk after 50 steps for different choices of the initial state. In the upper plot we choose the initial state $\psi_S=\frac{1}{2}(1,i,i,-1)^T$ which leads to a symmetric probability distribution with a dominant central spike. However, if we chose the initial state $\psi_G$ according to Eq. (\ref{grover:nospike:state}) we find that the central spike vanishes and most of the probability is situated at the edges, as depicted in the lower plot.}
\label{g3d}
\end{center}
\end{figure}

We start with the 2-D Grover walk which is driven by the coin
\begin{equation}
G = \frac{1}{2}\left(
                 \begin{array}{rrrr}
                   -1 & 1 & 1 & 1 \\
                   1 & -1 & 1 & 1 \\
                   1 & 1 & -1 & 1 \\
                   1 & 1 & 1 & -1 \\
                 \end{array}
               \right).
\label{grover:coin}
\end{equation}
It was identified numerically \cite{2dw1} and later proven analytically \cite{localisation} that the Grover walk exhibits a localisation effect, i.e. the probability $p_0(t)$ does not vanish but converges to a non-zero value except for a particular initial state
\begin{equation}
\psi_G\equiv\psi_G(0,0,0) = \frac{1}{2}\left(1,-1,-1,1\right)^T.
\label{grover:nospike:state}
\end{equation}
To illustrate this fact we present in \fig{g3d} the probability distribution of the Grover walk for different choices of the initial coin state, namely for $\psi_S=\frac{1}{2}(1,i,i,-1)^T$ and for $\psi_G$ given by Eq. (\ref{grover:nospike:state}).


In order to explain the localisation we analyze the eigenvalues of the time evolution operator in the Fourier picture defined by Eq. (\ref{teopF}) for the Grover walk
\begin{equation}
\widetilde{U}_G(k_1,k_2) = \left(D(k_1)\otimes D(k_2)\right) G.
\label{gkl}
\end{equation}
We find that they are given by
\begin{equation}
\label{eigenval:Grover}
\lambda_{1,2}  =  \pm 1,\qquad \lambda_{3,4}(k_1,k_2) = e^{\pm i \omega(k_1,k_2)}
\end{equation}
where the phase $\omega(k_1,k_2)$ reads
\begin{equation}
\cos(\omega(k_1,k_2)) = -\cos{k_1}\cos{k_2}.
\label{phase:Grover}
\end{equation}
The eigenvalues $\lambda_{1,2}$ are constant. As a consequence the return probability is non-vanishing as discussed in detail in Section~\ref{sec33}, unless the initial state is orthogonal to the eigenvectors corresponding to $\lambda_{1,2}$ at every point $(k_1,k_2)$. By explicitly calculating the eigenvectors of the matrix $\widetilde{U}_G(k_1,k_2)$ it is straightforward to see that such a vector is unique and equals that in Eq. (\ref{grover:nospike:state}), in agreement with the result derived in \cite{localisation}.

It is now easy to show that for the particular initial state given by Eq. (\ref{grover:nospike:state}) the probability $p_0(t)$ decays like $t^{-2}$. Indeed, as the initial state of Eq. (\ref{grover:nospike:state}) is orthogonal to the eigenvectors corresponding to $\lambda_{1,2}$ the asymptotic behaviour is determined by the remaining eigenvalues $\lambda_{3,4}(k_1,k_2)$, or more precisely by the saddle points of the phase $\omega(k_1,k_2)$. From Eq. (\ref{phase:Grover}) we find that it has only non-degenerate saddle points $k_1^0,\ k_2^0=\pm \pi/2$. For the initial state of Eq. (\ref{grover:nospike:state}) the probability that the Grover walk returns to the origin decays like $t^{-2}$. We conclude that the Grover walk on a 2-D lattice is recurrent and its P\'olya number equals one for all initial states except the one given in Eq. (\ref{grover:nospike:state}) for which the walk is transient.

\begin{figure}
\begin{center}
\includegraphics[width=0.4\textwidth]{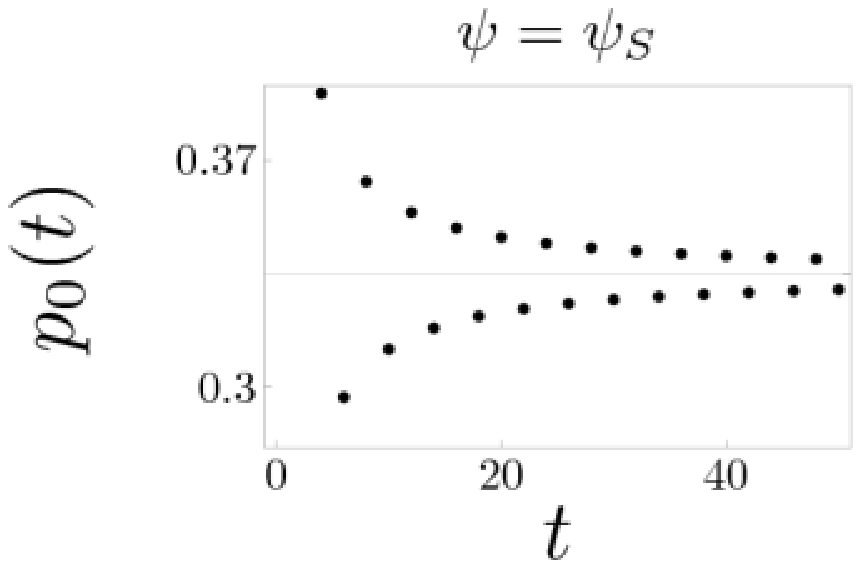}\vspace{12pt}
\includegraphics[width=0.4\textwidth]{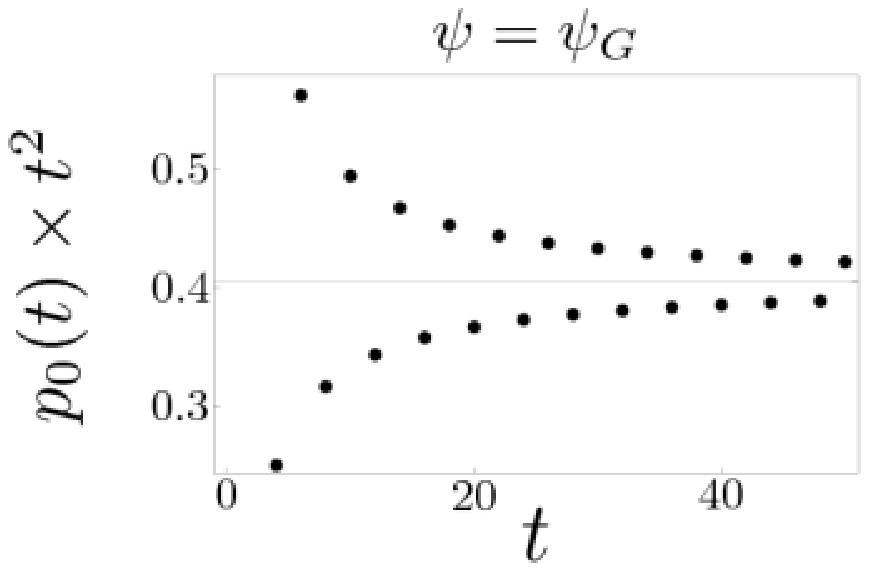}
\caption{The probability $p_0(t)$ for the Grover walk and different choices of the initial coin state. First, the walk starts with the coin state $\psi_S=\frac{1}{2}(1,i,i,-1)^T$ which leads to the non-vanishing value of $p_0(t)$, as depicted in the upper plot. The situation is the same for all initial coin states, up to the asymptotic value of $p_0(t\rightarrow +\infty)$, except for $\psi_G=\frac{1}{2}\left(1,-1,-1,1\right)^T$, as shown in the lower plot. Here we plot the probability $p_0(t)$ multiplied by $t^2$ to unravel the asymptotic behavior of $p_0(t)$. The plot confirms the analytic result of the scaling $p_0(t)\sim t^{-2}$.}
\label{gp0}
\end{center}
\end{figure}

We illustrate this result in \fig{gp0} where we plot the probability $p_0(t)$ for the two different choices of the initial coin states, namely $\psi_S=\frac{1}{2}(1,i,i,-1)^T$ and $\psi_G$. The plots confirm the analytical results of the scaling of the probability $p_0(t)$: on the upper plot we observe that $p_0(t)$ for the state $\psi_S$ oscillates around a nonzero value and thus has a non-vanishing limit, whereas on the lower plot we find that the probability $p_0(t)$ for the state $\psi_G$ decays like $t^{-2}$.


Let us estimate the P\'olya number of the Grover walk for the initial state of Eq. (\ref{grover:nospike:state}). The numerical simulations indicate that the return probability $p_0(t)$ for the initial state $\psi_G$ is the same as the return probability of the 2-D QW with tensor product coin. Hence, their P\'olya numbers coincide. With the help of the relation (\ref{Polya:ind:est}) we can estimate the P\'olya number of the Grover walk with the initial state of Eq. (\ref{grover:nospike:state}) by
\begin{equation}
P_G(\psi_G) \equiv P^{(2)}\approx 0.27325.
\end{equation}

\subsection{Recurrent quantum walks in arbitrary dimensions}
\label{sec52}

The above derived results allow us to construct for an arbitrary dimension $d$ a QW which is recurrent, except for a subspace of initial states.
Let us first consider the case when the dimension of the walk is even and equals $2d$. We choose the coin as a tensor product
\begin{equation}
G^{(2d)} = \otimes^d G
\label{c2d}
\end{equation}
of $d$ Grover coins given by Eq. (\ref{grover:coin}). As follows from Eqs. (\ref{teopF}) and (\ref{dk2}) the time evolution operator in the Fourier picture is also a tensor product
\begin{equation}
\widetilde{U}_G^{(2d)}(\textbf{k}) = \widetilde{U}_G(k_1,k_2)\otimes\ldots\otimes \widetilde{U}_G(k_{2d-1},k_{2d})
\label{c2d:k}
\end{equation}
of the matrices $\widetilde{U}_G$ defined by Eq. (\ref{gkl}) with different Fourier variables $k_i$. Hence, the eigenvalues of $\widetilde{U}_G^{(2d)}(\textbf{k})$ are given by the product of the eigenvalues of $\widetilde{U}_G$. Since two eigenvalues of $\widetilde{U}_G$ are constant as we have found in Eq. (\ref{eigenval:Grover}) one half of the eigenvalues of $\widetilde{U}_G^{(2d)}(\textbf{k})$ are also independent of $\textbf{k}$. As we have discussed in Section~\ref{sec33} the probability $p_0(t)$ converges to a non-zero value and therefore the QW exhibits localisation.

In the case of odd dimension $2d+1$ we augment the coin given by Eq. (\ref{c2d}) by the Hadamard coin for the extra spatial dimension
\begin{equation}
G^{(2d+1)} = G^{(2d)}\otimes C(0,0).
\label{c2d1}
\end{equation}
Performing a similar analysis as in the case of even dimensions we find that for the QW driven by the coin $G^{(2d+1)}$ the probability that the walk returns to the origin decays like $t^{-1}$ due to the Hadamard walk in the extra spatial dimension. Hence, this QW is recurrent.

We note that due the fact that the 2-D Grover walk is transient for the initial state $\psi_G$ the same statement holds for the above constructed QWs, supposed that the initial state contains $\psi_G$ in its tensor product decomposition. Such vectors form a subspace with dimension  equal $4^{d-1}$ for even dimensional walk and $2\times 4^{d-1}$ for odd dimensional walk.


\section{Recurrence of the 2-D Fourier walk}
\label{sec6}

We now turn to the 2-D Fourier walk driven by the coin
\begin{equation}
F = \frac{1}{2}\left(
                 \begin{array}{rrrr}
                   1 & 1 & 1 & 1 \\
                   1 & i & -1 & -i \\
                   1 & -1 & 1 & -1 \\
                   1 & -i & -1 & i \\
                 \end{array}
               \right).
\end{equation}
As we will see, the Fourier walk does not exhibit localisation. However, the decay of the probability $p_0(t)$ is slowed down to $t^{-1}$ so the Fourier walk is recurrent, except for a subspace of states.

We start our analysis of the Fourier walk with the matrix
\begin{equation}
\widetilde{U}_F(k_1,k_2) = \left(D(k_1)\otimes D(k_2)\right) F,
\label{fkl}
\end{equation}
which determines the time evolution in the Fourier picture. It seems to be hard to determine the eigenvalues of $\widetilde{U}_F(k_1,k_2)$ analytically. However, we only need to determine the saddle points of their phases $\omega_j(k_1,k_2)$. For this purpose we consider the eigenvalue equation
\begin{equation}
\Phi(k_1,k_2,\omega)\equiv\det{\left(\widetilde{U}_F(k_1,k_2)-e^{i \omega} I\right)}=0.
\end{equation}
This equation gives us the phases $\omega_i(k_1,k_2)$ as the solutions of the implicit function
\begin{eqnarray}
\nonumber \Phi(k_1,k_2,\omega) = & 1+\cos(2k_2)-2\cos(2\omega)+2\sin{2\omega}+&\\
\nonumber &   +4\cos{k_2}\sin{\omega}\left(\sin{k_1}-\cos{k_1}\right) = 0 & .\\
\label{fourier:implicit}
\end{eqnarray}
Using the implicit differentiation we find the derivatives of the phase $\omega$
\begin{eqnarray}
\nonumber \frac{\partial \omega}{\partial k_1} & = & -\frac{\cos{k_2}\sin{\omega}\left(\cos{k_1}+\sin{k_1}\right)}{\cos(2\omega)+\sin(2\omega)+\cos{k_2}\cos{\omega}\left(\sin{k_1}-\cos{k_2}\right)}\\
\nonumber \frac{\partial \omega}{\partial k_2} & = & -\frac{2\sin{k_2}\sin{\omega}\left(\cos{k_1}-\sin{k_1}\right)-\sin(2k_2)}{2\left(\cos(2\omega)+\sin(2\omega)+\cos{k_2}\cos{\omega}\left(\sin{k_1}-\cos{k_2}\right)\right)}\\
\label{fourier:der}
\end{eqnarray}
with respect to $k_1$ and $k_2$. Though we cannot eliminate $\omega$ on the RHS of Eq. (\ref{fourier:der}), we can identify the stationary points $\textbf{k}^0=(k_1^0,k_2^0)$
\begin{equation}
\left.\frac{\partial\omega(\textbf{k})}{\partial k_i}\right|_{\textbf{k}=\textbf{k}^0}=0,\quad i=1,2
\end{equation}
of $\omega(k_1,k_2)$ with the help of the implicit function $\Phi(k_1,k_2,\omega)$. We find the following:\\
({\it i}) $\omega_{1,2}(k_1,k_2)$ have saddle lines $$\gamma_1=(k_1,0)\ \textrm{and}\ \gamma_2=(k_1,\pi)$$\\
({\it ii}) all four phases $\omega_{i}(k_1,k_2)$ have saddle points for $$k_1^0=\frac{\pi}{4},\ -\frac{3\pi}{4}\quad \textrm{and}\quad k_2^0=\pm\frac{\pi}{2}$$

It follows from the case ({\it iii}) of Section~\ref{sec33} that the two phases $\omega_{1,2}(k_1,k_2)$ with saddle lines $\gamma_{1,2}$ are responsible for the slow down of the decay of the probability $p_0(t)$ to $t^{-1}$ for the Fourier walk, unless the initial coin state is orthogonal to the corresponding eigenvectors $v_{1,2}(k_1,k_2)$ at the saddle lines. For such an initial state the probability $p_0(t)$ behaves like $t^{-2}$ as the asymptotics of the integral given by Eq. (\ref{psi:0}) is determined only by the saddle points ({\it ii}).

Let us now determine the states $\psi_F$ which lead to the fast decay $t^{-2}$ of the probability that the Fourier walk returns to the origin. The states $\psi_F$ have to be constant vectors fulfilling the conditions
\begin{equation}
\left(\psi_F,v_{1,2}(\mathbf{k})\right)=0 \quad \forall\ \mathbf{k}\in\gamma_{1,2},
\end{equation}
which implies that $\psi_F$ must be a linear combination of $v_{3,4}(\mathbf{k}\in\gamma_{1,2})$ forming a two-dimensional subspace in $\mathcal{H}_C$. For $k_2=0,\pi$ we can find the eigenvectors of the matrix $\widetilde{U}_F(k_1,k_2)$ explicitly
\begin{eqnarray}
\nonumber v_1(k_1,0) & = & v_2(k_1,\pi) = \frac{1}{2}\left(e^{-ik_1},1,-e^{-ik_1},1\right)^T\\
\nonumber v_1(k_1,\pi) & = & v_2(k_1,0) = \frac{1}{2}\left(-e^{-ik_1},1,e^{-ik_1},1\right)^T\\
\nonumber v_3(k_1,0) & = & v_3(k_1,\pi) = \frac{1}{\sqrt{2}}(1,0,1,0)^T \\
v_4(k_1,0) & = & v_4(k_1,\pi) = \frac{1}{\sqrt{2}}(0,1,0,-1)^T.
\end{eqnarray}
The explicit form of $\psi_F$ reads
\begin{equation}
\psi_F(a,b) = \left(a,b,a,-b\right)^T,
\label{psi:F}
\end{equation}
where $a,b\in \mathds{C}$. We point out that the particular initial state
\begin{equation}
\psi_F\left(a=\frac{1}{2},b=\frac{1-i}{2\sqrt{2}}\right) = \frac{1}{2}\left(1,\frac{1-i}{\sqrt{2}},1,-\frac{1-i}{\sqrt{2}}\right)^T
\label{sym:F}
\end{equation}
which was identified in \cite{2dw1} as the state which leads to a symmetric probability distribution with no peak in the neighborhood of the origin belongs to the family described by Eq. (\ref{psi:F}).

\begin{figure}
\begin{center}
\includegraphics[width=0.45\textwidth]{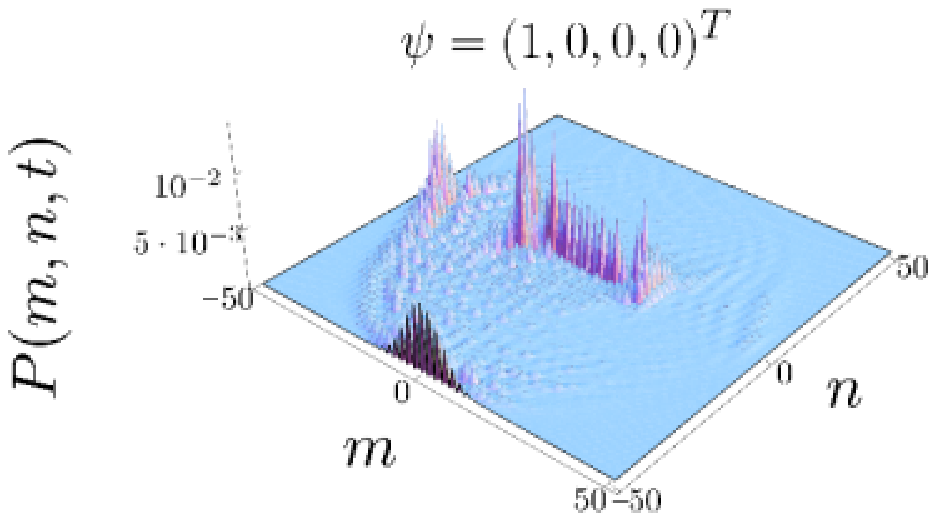}\vspace{12pt}
\includegraphics[width=0.35\textwidth]{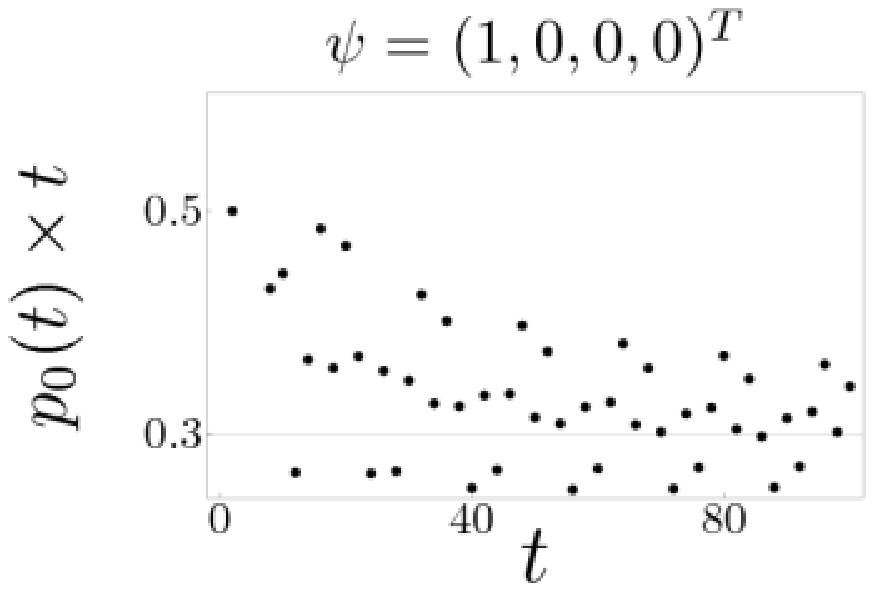}
\caption{Probability distribution after 50 steps and the time evolution of the probability $p_0(t)$ for the Fourier walk with the initial state $\psi=(1,0,0,0)^T$. The upper plot of the probability distribution reveals a presence of the central peak. Indeed, $\psi$ is not a member of the family $\psi_F(a,b)$. However, in contrast to the Grover walk the peak decays vanishes which. In the lower plot we illustrate this by showing the probability $p_0(t)$ multiplied by $t$ to unravel the asymptotic behaviour $p_0(t)\sim t^{-1}$.}
\label{f3d1}
\end{center}
\end{figure}

\begin{figure}
\begin{center}
\includegraphics[width=0.45\textwidth]{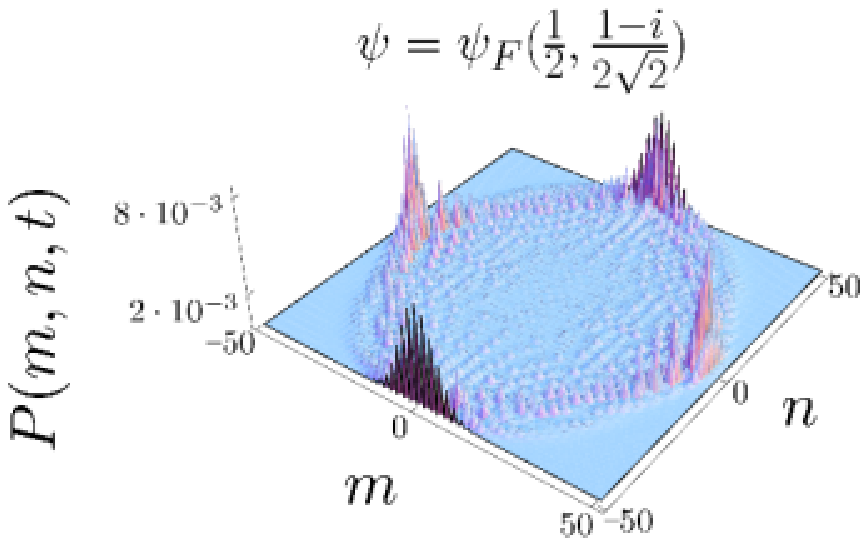}\vspace{12pt}
\includegraphics[width=0.35\textwidth]{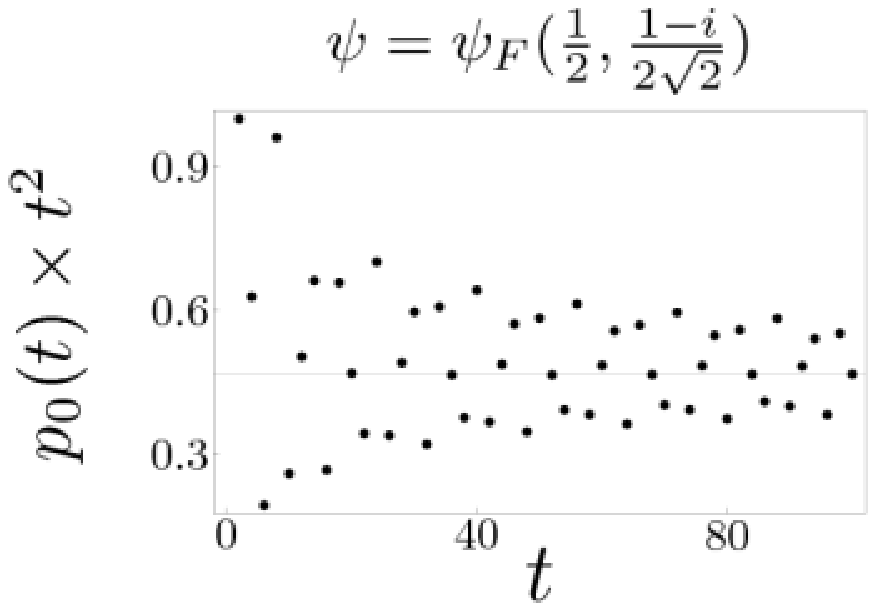}
\caption{Probability distribution after 50 steps and the time evolution of the probability $p_0(t)$ for the Fourier walk with the initial state given by Eq. (\ref{sym:F}). Since $\psi$ is a member of the family $\psi_F(a,b)$ the central peak in the probability distribution is not present, as depicted on the upper plot. The lower plot indicates that the probability $p_0(t)$ decays like $t^{-2}$.}
\label{f3d2}
\end{center}
\end{figure}

We illustrate the results in \fig{f3d1} and \fig{f3d2}. In \fig{f3d1} we plot the probability distribution and the probability $p_0(t)$ for the Fourier walk with the initial state $\psi=(1,0,0,0)^T$. This vector is not a member of the family $\psi_F(a,b)$ defined by Eq. (\ref{psi:F}). We find that a central peak is present, as depicted on the upper plot. However, in contrast to the Grover walk, the peak vanishes as shown on the lower plot, where we plot the probability $p_0(t)$ multiplied by $t$. Nevertheless, the plot indicates that the probability $p_0(t)$ decays like $t^{-1}$, in agreement with the analytical result. In contrast, for \fig{f3d2} we have chosen the initial state given by Eq. (\ref{sym:F}) which is a member of the family $\psi_F(a,b)$. The upper plot shows highly symmetric probability distribution. However, the central peak is not present and as the lower plot indicates the probability $p_0(t)$ decays like $t^{-2}$.


We conclude that the Fourier walk is recurrent except for the two-dimensional subspace of initial states defined by Eq. (\ref{psi:F}) for which the walk is transient.

Let us now turn to the estimation of the P\'olya numbers of the 2-D Fourier walk for the two-dimensional subspace of initial states given by Eq. (\ref{psi:F}). We make use of the normalization condition and the fact that the global phase of a state is irrelevant. Hence, we can choose $a$ to be non-negative real and $b$ is then given by the relation
\begin{equation}
b=\sqrt{\frac{1}{2}-a^2}e^{i\phi}.
\end{equation}
Therefore, we parameterize the family of states defined by Eq. (\ref{psi:F}) by two real parameters --- $a$ ranging from $0$ to $\frac{1}{\sqrt{2}}$ and the mutual phase $\phi\in[0,2\pi)$. The exact expression for $p_0(a,\phi,t)$ can be written in the form
\begin{equation}
p_0(a,\phi,t)=\frac{K_1(t)-K_2(t) a\sqrt{\frac{1}{2}-a^2}(\cos{\phi}-\sin{\phi})}{t^2},
\end{equation}
where $K_{1,2}$ has to be determined numerically. Nevertheless, the numerical simulation of $p_0(a,\phi,t)$ at two values of $(a,\phi)$ enables us to find the numerical values of $K_{1,2}(t)$ and we can evaluate $p_0(a,\phi,t)$ at any point $(a,\phi)$. The probability $p_0(a,\phi,t)$ shows the maximum at $a=\frac{1}{2}$, $\phi=\frac{3\pi}{4}$ and the minimum for the same value of $a$ and the phase $\phi=\frac{7\pi}{4}$. Consequently, these points also represent the maximum and the minimum of the P\'olya numbers.


\begin{figure}[h]
\begin{center}
\includegraphics[width=0.45\textwidth]{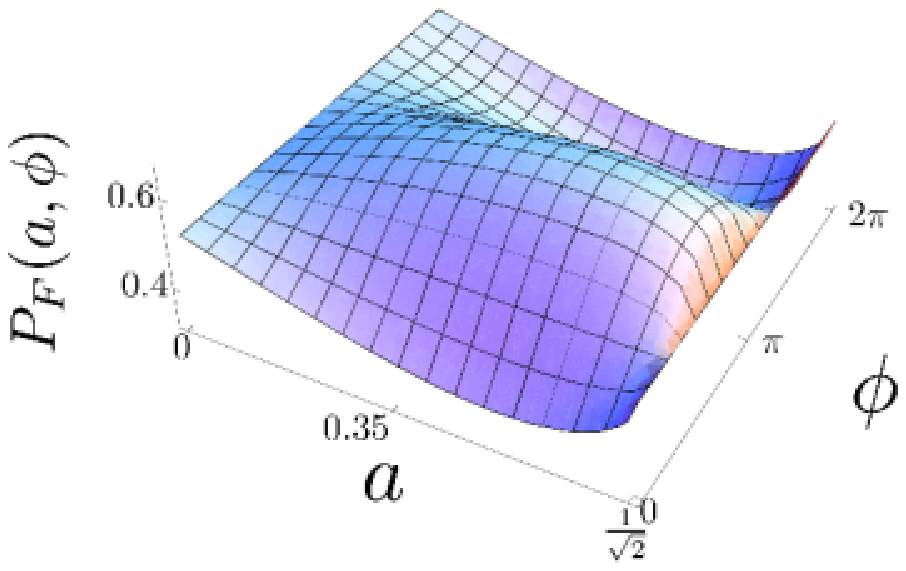}\vspace{12pt}
\includegraphics[width=0.45\textwidth]{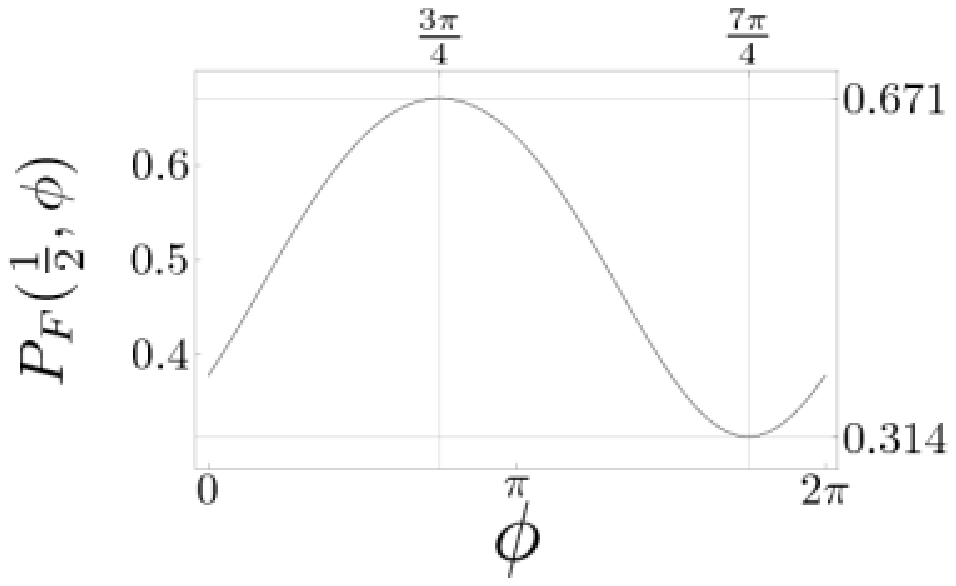}
\caption{Approximation of the P\'olya numbers for the 2-D Fourier walk and the initial states from the family of states defined by Eq. (\ref{psi:F}) in their dependence on the parameters of the initial state $a$ and $\phi$. Here we have evaluated the first 100 terms of $p_0(a,\phi,t)$ exactly. The P\'olya numbers cover the whole interval between the minimal value of $P_F^{min}\approx 0.314$ and the maximal value of $P_F^{max}\approx 0.671$. The extreme values are attained for $a=1/2$ and $\phi^{min}=7\pi/4$, respectively $\phi^{max}=3\pi/4$. On the lower plot we show the cut at the value $a=1/2$ containing both the maximum and the minimum.}
\label{polya:fourier}
\end{center}
\end{figure}


In \fig{polya:fourier} we present the approximation of the P\'olya number Eq. (\ref{polya:approx}) in its dependence on $a$ and $\phi$ and a cut through the plot at the value $a=1/2$. Here we have evaluated the first 100 terms of $p_0(a,\phi,t)$ exactly. We see that the values of the P\'olya number vary from the minimum $P_F^{min}\approx 0.314$ to the maximal value of $P_F^{max}\approx 0.671$. We note that for the initial states that do not belong to the subspace defined by Eq. (\ref{psi:F}) the P\'olya number equals one.


\section{Conclusions}
\label{sec7}

Our results, summarized in \tab{tab2}, demonstrate that there is a remarkable freedom for the value of the P\'olya number in higher dimensions, depending both on the initial state and the coin operator, in contrast to the classical random walk where the dimension of the lattice uniquely defines the recurrence probability. Hence, the quantum P\'olya number is able to indicate physically different regimes in which a QW can be operated in. We expect further interesting effects when we relax the condition of allowing only for unit steps and introduce larger jumps. In that case the size of the coin operator can exceed the dimension of the lattice thus in low dimensional lattices some effects seen in higher dimensions can be anticipated, e.g. localisation has been found in three-state quantum walks on a 1-D lattice \cite{1dloc}.

\begin{table}
\begin{center}
\begin{tabular}{|c|c|c|c|}
  \hline
  \multirow{2}{*}{QW} & \multirow{2}{*}{Section} & \multirow{2}{*}{Return probability} & \multirow{2}{*}{P\'olya number}\\
  & & & \\\hline
  \multirow{2}{*}{1-D unbiased}  & \multirow{2}{*}{\ref{sec41}} & \multirow{2}{*}{$t^{-1}$ for any $\psi$} & \multirow{2}{*}{1} \\
  & & & \\\hline
  \multirow{2}{*}{$d$-D TP coin} & \multirow{2}{*}{\ref{sec42}} & \multirow{2}{*}{$t^{-d}$ for any $\psi$} & $<1$\\
  & & & independent of $\psi$ \\\hline
  \multirow{4}{*}{2-D Grover} & \multirow{4}{*}{\ref{sec51}}  & \multirow{2}{*}{const. for $\psi\neq\psi_G$} & \multirow{2}{*}{1}\\
  & & & \\
  & & \multirow{2}{*}{$t^{-2}$ for $\psi=\psi_G$} & \multirow{2}{*}{$<1$}\\
  & & & \\\hline
  \multirow{4}{*}{2-D Fourier} & \multirow{4}{*}{\ref{sec6}}  & \multirow{2}{*}{$t^{-1}$ for $\psi\not\in\psi_F$} & \multirow{2}{*}{1}\\
  & & & \\
  & & \multirow{2}{*}{$t^{-2}$ for $\psi\in\psi_F$} & $<1$\\
  & & & dependent on $\psi$ \\\hline
\end{tabular}
\caption{Summary of the main results. We list the types of studied QWs, the asymptotic behaviour of the return probability and the P\'olya number in the respective cases in its dependence on the initial state $\psi$.}
\label{tab2}
\end{center}
\end{table}

In the present paper we assumed a specific measurement scheme where the dynamics are not continued after the measurement is performed. We note that this is only one of the possibilities to define the P\'olya number, one could vary the frequency of measurements randomly or in a deterministic manner while continuing the time evolution. The present definition has the advantage of maintaining unitary time evolution, thus a pure state for initial pure states.


\begin{acknowledgments}

We thank W. P. Schleich and J. Asb\'oth for stimulating discussions. The financial support by MSM 6840770039, M\v SMT LC 06002, the Czech-Hungarian cooperation project (KONTAKT,CZ-2/2008) and by the Hungarian Scientific Research Fund (T049234) is gratefully acknowledged.

\end{acknowledgments}


\appendix

\section{Recurrence criterion for quantum walks}
\label{app:a}

In this Appendix we show that the recurrence criterion for QWs is the same as for RWs, i.e. the P\'olya number equals one if and only if the series
\begin{equation}
{\cal S} \equiv \sum_{t=0}^{\infty}p_0(t)
\end{equation}
diverges.

According to the definition of the P\'olya number Eq. (\ref{polya:def}) for QWs we have to prove the equivalence
\begin{equation}
\overline{P}\equiv\prod\limits_{t=1}^{+\infty}\left(1-p_0(t)\right) = 0 \Longleftrightarrow {\cal S}=+\infty.
\end{equation}
We note that the convergence of both the sum ${\cal S}$ and the product $\overline{P}$ is unaffected if we omit finitely many terms.

Let us first consider the case when the sequence $p_0(t)$ converges to a non-zero value $0<a\leq 1$. Obviously, in such a case the series ${\cal S}$ is divergent. Since $p_0(t)$ converges to $a$ we can find for any $\varepsilon>0$ some $t_0$ such that for all $t>t_0$ the inequalities
\begin{equation}
1-a-\varepsilon\leq 1-p_0(t)\leq 1-a+\varepsilon.
\end{equation}
hold. Hence, we can bound the infinite product
\begin{equation}
\lim\limits_{t\rightarrow +\infty}\left(1-a-\varepsilon\right)^t\leq\overline{P}\leq\lim\limits_{t\rightarrow +\infty} \left(1-a+\varepsilon\right)^t.
\label{app1}
\end{equation}
Since we can choose $\varepsilon$ such that
\begin{equation}
\left|1-a\pm\varepsilon\right|<1,
\end{equation}
we find that limits both on the left-hand side and the right-hand side of Eq. (\ref{app1}) equals zero. Hence, the product $\overline{P}$ vanishes.

Let us now turn to the case when $p_0(t)$ converges to zero. We denote the partial product
\begin{equation}
\overline{P}_n = \prod\limits_{t=1}^n(1-p_0(t)).
\end{equation}
Since $1-p_0(t)>0$ for all $t\geq 1$ we can consider the logarithm
\begin{equation}
\ln{\overline{P}_n} = \sum\limits_{t=1}^n\ln\left(1-p_0(t)\right)
\label{app3}
\end{equation}
and rewrite the infinite product as a limit
\begin{equation}
\overline{P} = \lim\limits_{n\rightarrow +\infty}e^{\ln{\overline{P}_n}}
\label{app2}.
\end{equation}
Since $p_0(t)$ converges to zero we can find some $t_0$ such that for all $t>t_0$ the value of $p_0(t)$ is less or equal than $1/2$. With the help of the inequality
\begin{equation}
-2x\leq\ln\left(1-x\right)\leq -x
\end{equation}
valid for $x\in\left[0,1/2\right]$ we find the following bounds
\begin{equation}
-2\sum\limits_{t=1}^n p_0(t)\leq\ln\overline{P}_n\leq -\sum_{t=1}^n p_0(t).
\end{equation}
Hence, if the series ${\cal S}$ is divergent the limit of the sequence ${\left(\ln{\overline{P}_n}\right)}^\infty_{n=1}$ is $-\infty$ and according to Eq. (\ref{app2}) the product $\overline{P}$ vanishes. If, on the other hand, the series ${\cal S}$ converges the sequence ${\left(\ln{\overline{P}_n}\right)}^\infty_{n=1}$ is bounded. According to Eq. (\ref{app3}) the partial sums of the series $\sum\limits_{t=1}^{+\infty}\ln\left(1-p_0(t)\right)$ are bounded and since it is a series with strictly negative terms it converges to some negative value $b<0$. Consequently, the sequence ${\left(\ln{\overline{P}_n}\right)}^\infty_{n=1}$ converges to $b$ and according to Eq. (\ref{app2}) the product equals
\begin{equation}
\overline{P} = e^b>0.
\end{equation}
This completes our proof.



\begin{thebibliography}{x}
\bibitem{overview}
see e.g. N. Guillotin-Plantard and R. Schott, {\it Dynamic Random Walks: Theory and Application}, Elsevier, Amsterdam (2006)
\bibitem{rw:compsc1}
C. Papadimitriou, {\it Computational Complexity}, Addison Wesley, Reading (1994)
\bibitem{rw:compsc2}
R. Motwani and P. Raghavan, {\it Randomized Algorithms}, Cambridge University Press, Cambridge (1995)
\bibitem{graph:connect}
see e.g. A. Sinclair, {\it Algorithms for Random Generation and Counting, a Markov Chain Approach}, Birkhauser Press, Boston (1993)
\bibitem{3-sat}
U. Sch\"oning, 40th Annual Symposium on Foundations of Computer Science, IEEE, New York, 17 (1999)
\bibitem{matrix:perm}
M. Jerrum, A. Sinclair and E. Vigoda, in Proceedings of the 33th STOC, New York, 712 (2001)
\bibitem{aharonov}
Y. Aharonov, L. Davidovich and N. Zagury, Phys. Rev. A \textbf{48}, 1687 (1993)
\bibitem{meyer1}
D. Meyer, J. Stat. Phys. \textbf{85}, 551 (1996)
\bibitem{meyer2}
D. Meyer, Phys. Lett. A \textbf{223}, 337 (1996)
\bibitem{watrous}
J. Watrous, J. Comput. Syst. Sci. \textbf{62}, 376 (2001)
\bibitem{farhi}
E. Farhi and S Gutmann, Phys. Rev. A \textbf{58}, 915 (1998)
\bibitem{childs}
A. Childs, E. Farhi and S. Gutmann, Quantum Inf. Process. \textbf{1}, 35 (2002)
\bibitem{hillery:2003}
M. Hillery, J. Bergou and E. Feldman, Phys. Rev. A \textbf{68}, 032314 (2003)
\bibitem{hillery:2004}
E. Feldman, M. Hillery, Phys. Lett. A \textbf{324}, 277 (2004)
\bibitem{kosik:2005}
J. Ko\v s\'{\i}k and V. Bu\v zek, Phys. Rev. A \textbf{71}, 012306 (2005)
\bibitem{hillery:2007}
E. Feldman and M. Hillery, J. Phys. A \textbf{40}, 11343 (2007)
\bibitem{strauch}
F. W. Strauch, Phys. Rev. A \textbf{74}, 030301 (2006)
\bibitem{kempe}
D. Aharonov, A. Ambainis, J. Kempe and U. Vazirani, in Proceedings of the 33th STOC, New York, 50 (2001)
\bibitem{ambainis}
A. Ambainis, E. Bach, A. Nayak, A. Vishwanath and J. Watrous, Proceedings of the 33th STOC, New York, 60 (2001)
\bibitem{shenvi:2003}
N. Shenvi, J. Kempe and K. B. Whaley, Phys. Rev. A \textbf{67}, 052307 (2003)
\bibitem{ambainis:2003}
A. Ambainis, arXiv:quant-ph/0311001
\bibitem{kendon:2006}
V. Kendon, Phil. Trans. R. Soc. A 364, 3407 (2006)
\bibitem{magniez}
F. Magniez, A. Nayak, J. Roland and M. Santha, arXiv:quant-ph/0608026
\bibitem{aurel:2007}
A. Gabris, T. Kiss and I. Jex, Phys. Rev. A \textbf{76}, 062315 (2007)
\bibitem{2dw1}
B. Tregenna, W. Flanagan, R. Maile and V. Kendon, New J. Phys. \textbf{5}, 83.1 (2003)
\bibitem{miyazaki}
T. Miyazaki, M. Katori and N. Konno, Phys. Rev. A \textbf{76}, 012332 (2007)
\bibitem{chandrashekar:2007}
C.M. Chandrashekar, R. Srikanth and R. Laflamme, arXiv:0711.1882
\bibitem{bach:2004}
E. Bach, S. Coppersmith, M. P. Goldschen, R. Joynt and J. Watrous, J. Comput. Syst. Sci. \textbf{69}, 562 (2004)
\bibitem{kempe:2005}
J. Kempe, Prob. Th. Rel. Fields \textbf{133} (2), 215 (2005)
\bibitem{krovi:2006a}
H. Krovi and T. A. Brun, Phys. Rev. A \textbf{73}, 032341 (2006)
\bibitem{krovi:2006b}
H. Krovi and T. A. Brun, Phys. Rev. A \textbf{74}, 042334 (2006)
\bibitem{kendon:2006b}
V. Kendon, Math. Struct. in Comp. Sci \textbf{17}(6), 1169 (2006)
\bibitem{varbanov:2008}
M. Varbanov, H. Krovi and T. A. Brun, arXiv:0803.3446
\bibitem{nayak}
A. Nayak and A. Vishwanath, arXiv:quant-ph/0010117v1
\bibitem{konno:2002}
N. Konno, Quantum Inform. Compu. \textbf{2}, 578 (2002)
\bibitem{konno:2005b}
N. Konno, J. Math. Soc. Japan \textbf{57}, 1179 (2005)
\bibitem{carteret}
H. A. Carteret, M. E. H. Ismail and B. Richmond, J. Phys. A \textbf{36}, 8775 (2003)
\bibitem{Grimmett}
G. Grimmett, S. Janson and P. F. Scudo, Phys. Rev. E \textbf{69}, 026119 (2004)
\bibitem{2dqw}
T. D. Mackay, S. D. Bartlett, L. T. Stephenson and B. C. Sanders, J. Phys. A \textbf{35}, 2745 (2002)
\bibitem{localisation}
N. Inui, Y. Konishi and N. Konno, Phys. Rev. A \textbf{69}, 052323 (2004)
\bibitem{1dloc}
N. Inui, N. Konno and E. Segawa,  Phys. Rev. E \textbf{72}, 056112 (2005)
\bibitem{sato:2008}
M. Sato, N. Kobayashi, M. Katori and N. Konno, arXiv:0802.1997v1
\bibitem{sanders}
B. C. Sanders, S. D. Bartlett, B. Tregenna and P. L. Knight, Phys. Rev. A \textbf{67}, 042305 (2003)
\bibitem{jeong}
H. Jeong, M. Paternostro and M. S. Kim, Phys. Rev. A \textbf{69}, 012310 (2004)
\bibitem{pathak}
P. K. Pathak and G. S. Agarwal, Phys. Rev. A \textbf{75}, 032351 (2007)
\bibitem{dur}
W. D\"ur, R. Raussendorf, V.M. Kendon and H.-J. Briegel, Phys. Rev. A \textbf{66}, 052319 (2002)
\bibitem{eckert}
K. Eckert, J. Mompart, G. Birkl and M. Lewenstein, Phys. Rev. A \textbf{72}, 012327 (2005)
\bibitem{chandrashekar:2006}
C.M. Chandrashekar, Phys. Rev. A \textbf{74}, 032307 (2006)
\bibitem{bruss:leuchs}
D. Bru\ss\  and G. Leuchs (Eds.), {\it Lectures on Quantum Information}, Wiley-VCH, Berlin (2006)
\bibitem{polya}
G. P\'olya, Mathematische Annalen \textbf{84}, 149 (1921)
\bibitem{hughes}
B. D. Hughes, {\it Random walks and random environments, Vol. 1: Random walks}, Oxford University Press, Oxford (1995)
\bibitem{domb:1954}
C. Domb, Proc. Cambridge Philos. Soc. \textbf{50}, 586 (1954)
\bibitem{montroll:1964}
E.W. Montroll, in {\it Random Walks on Lattices}, edited by R. Bellman (American Mathematical Society, Providence, RI), Vol. \textbf{16}, 193 (1964)
\bibitem{yang:2007}
W. Yang, C. Liu and K. Zhang, J. Phys. A \textbf{40}, 8487 (2007)
\bibitem{prl}
M. \v Stefa\v n\'ak, I. Jex and T. Kiss, Phys. Rev. Lett. \textbf{100}, 020501 (2008)
\bibitem{revesz}
P. R\'ev\'esz, {\it Random walk in random and non-random environments}, World Scientific, Singapore (1990)
\bibitem{jarnik}
V. Jarn\'{\i}k, {\it Diferenci\'aln\'{\i} po\v cet II}, Academia, Prague, 121 (1976)
\bibitem{dita}
see e.g. P. Dita, J. Phys. A \textbf{37}, 5355 (2004)
\bibitem{statphase}
R. Wong, {\it Asymptotic Approximations of Integrals}, SIAM, Philadelphia (2001)
\end{thebibliography}
\end{document}